\documentclass[]{spie}  

\usepackage{amsmath,amsfonts,amssymb}
\usepackage{graphicx}
\usepackage[colorlinks=true, allcolors=blue]{hyperref}
\usepackage{color}

\title{Design and test of optical payload for polarization encoded QKD for Nanosatellites}

\author[a,b]{Jaya Sagar}
\author[a,b]{Elliott Hastings}
\author[b]{Piede Zhang}
\author[b]{Milan Stefko}
\author[b]{David Lowndes}
\author[c]{Daniel Oi}
\author[b]{John Rarity}
\author[b]{Siddarth K. Joshi}

\affil[a]{Quantum Engineering Centre for Doctoral Training, University of Bristol, BS8 1FD, United Kingdom}
\affil[b]{Quantum Engineering Technology Labs, NSQI, University of Bristol, BS8 1FD, United Kingdom}
\affil[c]{Computational Nonlinear and Quantum Optics (CNQO) group, University of Strathclyde, Glasgow, G4 0NG, United Kingdom}
\authorinfo{Further author information: (Send correspondence to J. Sagar)\\ E-mail: jaya.sagar@bristol.ac.uk}

\begin{document} 
\maketitle

\begin{abstract}
 Satellite based Quantum Key Distribution (QKD) in Low Earth Orbit (LEO) is currently the only viable technology to span thousands of kilometres. Since the typical overhead pass of a satellite lasts for a few minutes, it is crucial to increase the the signal rate to maximise the secret key length. For the QUARC CubeSat mission due to be launched within two years, we are designing a dual wavelength, weak-coherent-pulse decoy-state Bennett-Brassard '84 (WCP DS BB84) QKD source. The optical payload is designed in a $12{\times}9{\times}5 cm^3$ bespoke aluminium casing. The Discrete Variable QKD Source consists of two symmetric sources operating at 785 nm and 808 nm. The laser diodes are fixed to produce Horizontal,Vertical, Diagonal, and Anti-diagonal (H,V,D,A) polarisation respectively, which are combined and attenuated to a mean photon number of 0.3 and 0.5 photons/pulse. We ensure that the source is secure against most side channel attacks by spatially mode filtering the output beam and characterising their spectral and temporal characterstics. The extinction ratio of the source contributes to the intrinsic Qubit Error Rate(QBER) with  $0.817 \pm 0.001\%$.   This source operates at 200MHz, which is enough to provide secure key rates of a few kilo bits per second despite 40 dB of estimated loss in the free space channel \cite{sidhu2022finite}.
\end{abstract}


\section{INTRODUCTION}
\label{sec:intro}  
Quantum Communication is emerging as a feasible method of sharing a secure key between two parties~\cite{2020_Pirandola} based on the laws of physics. The secret information contained in the quantum state cannot be cloned or measured without detectable disturbance. This attribute creates a challenge of loss intolerance of Quantum Communication channels. State-of-the-art Quantum Key Distribution (QKD) protocols have been limited to 1000 km in optical fibres due to the exponential losses ~\cite{2016_Yin,2020_Chen,2020_Boaron}. It is not yet realistic to use quantum repeaters for global extension of quantum networks due to the exhaustive requirements of a stable fibre optic network~\cite{2021_Mustafa}. 

The comparatively lower loss media of free space or satellite communication can offer a more flexible, scalable and less expensive platform for communication systems~\cite{sidhu2021advances}. Various studies have shown the feasibility of satellite quantum communication extending thousands of kilometres~\cite{2002_Rarity, 2003_Aspelmeyer}. Satellite based quantum communication can join various fibre based quantum channels towards global network ~\cite{2021_Chen}. In 2017, the  Micius mission (weighing 635 kg) successfully demonstrated intercontinental QKD from the LEO ~\cite{2017_Liao}. The reliable communication between LEO satellite and the ground station can be established when the satellite is $10^{\circ} $ above the horizon in 8-10 minutes~\cite{sidhu2022finite}. This makes it important to achieve high data transmission rates to get a significant key. Thus, efforts are being made to optimise the connection and improve transmission rates. With current advancements in space technology, the access to space has been made a lot easier with miniature satellites, CubeSats (1U to 12U)~\footnote{1U is a standard dimension (Units or “U”) of approximately $10cm{\times}10cm{\times}10 cm$ in the CubeSat standard.}~\cite{2020_Villar} or the Canada developed microsatellite mission QEYSSat~\cite{CanadaWeb}. However, the restricted space, weight and Power (SWaP) pose a range of complications for payload design and integration.

This paper shows a design and test of optical payload for polarization encoded QKD for the Quantum Communication Hub mission~\cite{2020_Mazzarella}. The source has been designed to fit within a shared 2U payload volume on a 12U platform. .The source comprises of two symmetrical counterparts at different wavelengths but results from only one wavelength are shown for this paper. The first section shows the design and assembly of the source optical module. The second sections outlines the characterisation and compatibility tests on the source to perform QKD at 100MHz on each wavelength. 

\section{Source Design}
 
   \begin{figure} [!h]
   \begin{center}
   \begin{tabular}{c} 
   \includegraphics[width=\textwidth]{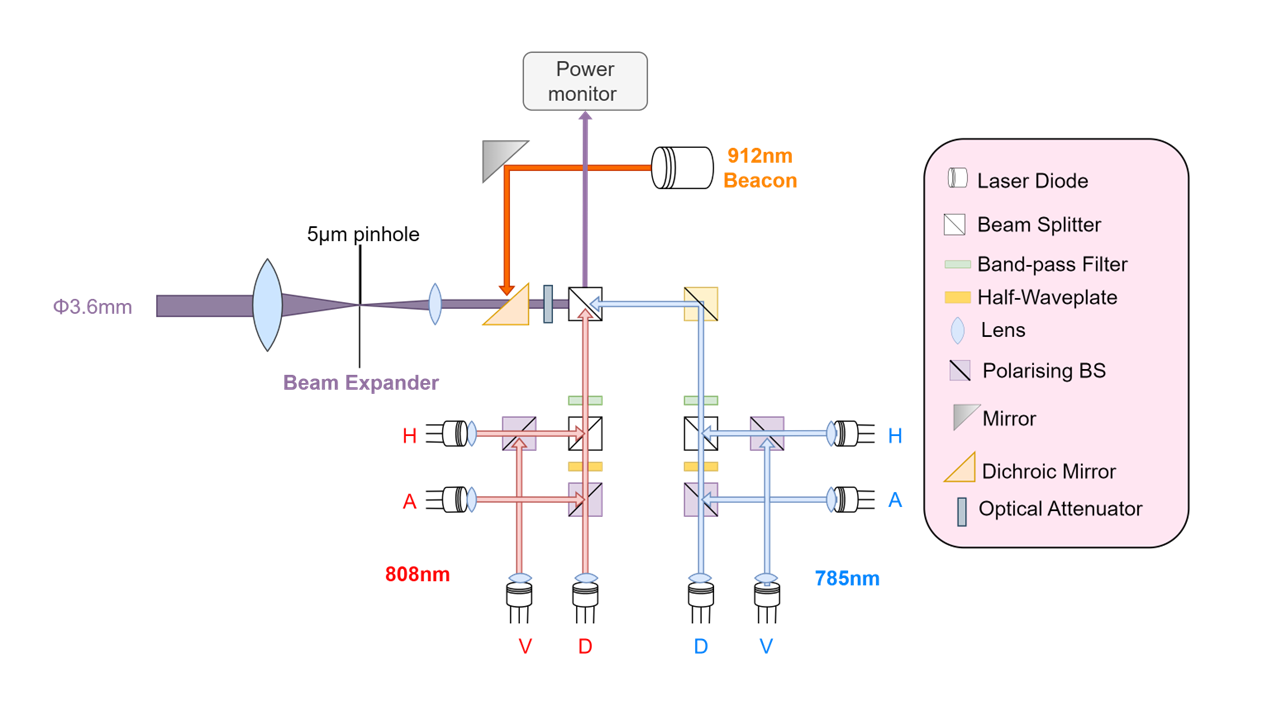}
	\end{tabular}
	\end{center}
   \caption[example] 
   { \label{fig:setup} 
 Schematic of the optical design of the source}
   \end{figure}


The optical payload is designed in a shared $20{\times}7{\times}5 cm^3$ aluminium frame. The actual size of  the Discrete Variable (DV) QKD source is  within $12\,cm \times 9\,cm \times 5cm $. It consists of two decoy state BB84 sources operating at different wavelengths. The protocol requires the source to encode information on the polarization of the photon, the receiver randomly chooses a basis to measure the polarization of the received photons and shares the basis chosen with the sender. If the sender and receiver used the same basis, the key is generated ~\cite{BB84}. This design as shown in \ref{fig:setup} consists of two BB84 sources operating at $785\, nm$ and $808\, nm$ wavelengths. This doubles the repetition rate of the pulses with limited electronic power on board. The dual wavelength source will provide research insights on QKD performance at different wavelengths. The set of laser diodes is selected from same batch and tested for indistinguishibility in Current-Voltage-Luminescence characteristics and spectrum. These are fixed orthogonal  to produce H and V polarisation respectively.The beams from the laser diodes are collimated using aspheric lenses. The customised zero order half wave plate set at $22.5^\circ$ is used to convert one pair of H,V to D,A polarisation.  The H,V,D,A are then combined using a Beam Splitter(BS). The $785\,nm$ beam is reflected to the output by a 10:90 BS.  The power from both the sources can be equalised using extra optical attenuators and  combined by a 50:50 beam splitter. One output of this beam splitter goes into the power monitor while the other goes into the beam expander via an optical attenuator to achieve the required mean photon number. The power monitor sets a reference for the expected power from the QKD source and when correlated to the inbuilt photodiode in the laser diodes package, ensures the health of the diodes and alignment for in-flight health checks. The attenuated output beams are coupled into a Keplerian Beam expander ~\footnote{A Keplerian beam expander has two converging lenses, separated by the sum of their focal lengths} to obtain a collimated $1.8\,mm$ at $1/e^2$ radius beam sent to the output optics for transmission to the optical ground station on Earth.

\begin{figure} [ht]
  \begin{center}
  \begin{tabular}{c} 
  \includegraphics[height=5cm]{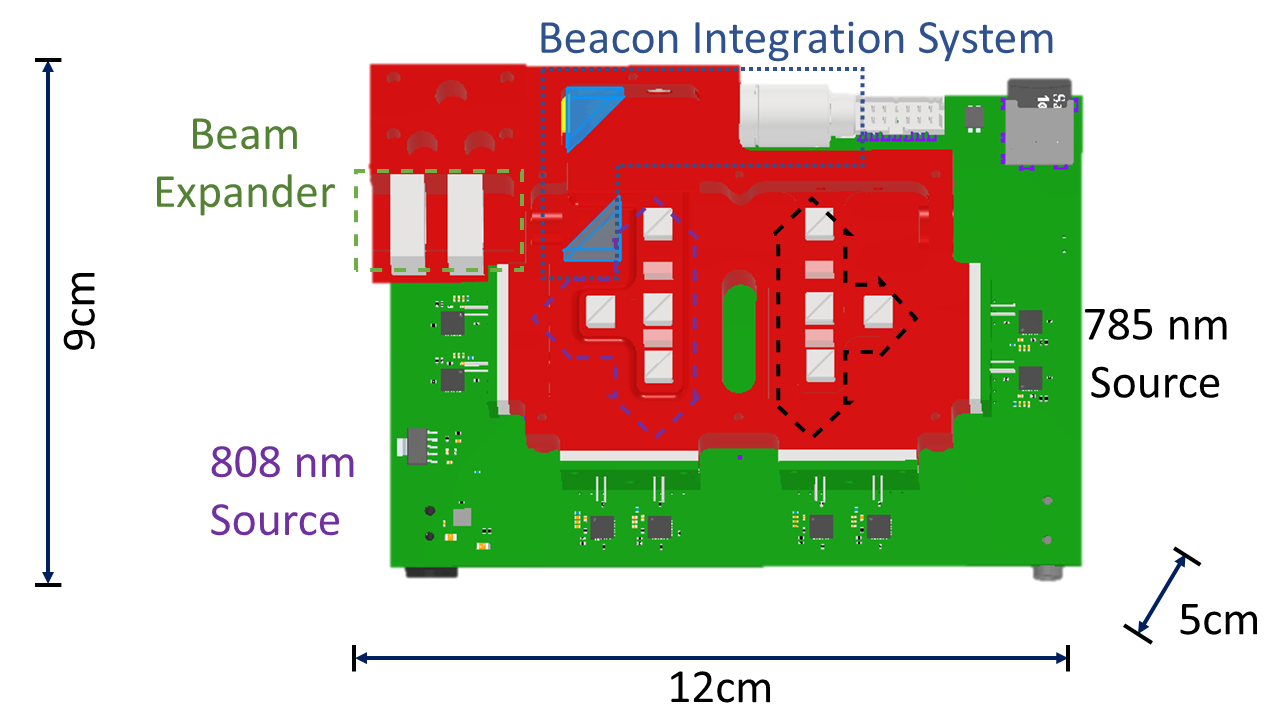}
	\end{tabular}
	\end{center}
  \caption[example] 
  { \label{fig:bespoke} 
Screenshot of the bespoke aluminium casing for the dual wavelength QKD source optical module.}
  \end{figure}

A bespoke aluminium body design (figure~\ref{fig:bespoke}) is machined to hold the optics through the temperature, pressure and vibration effects. To ensure the optical components are mounted stress-free, the optical components are glued with coefficient of thermal expansion (CTE) matched UV curing glue. The base thickness of $2\,mm$ is maintained all around the source to bring the radiation levels below $10\,krad$ in a 500 km sun-synchronous orbit orbit and prevent any direct radiation damage of the optics. This bespoke design is less than $100\,g$ and has slots to connect to the rest of the frame of the satellite and the electronics. It is anodised and acid polished $(10 \mu m)$ to prevent any short circuits with the electronics.  The total weight of the source module including all the optics and electronics is $\le300\,g$
 
The optical components in the source are aligned using a vacuum pick-up tool which is connected to a Smaract modular multidimensional positioning system. This gives us nanometre lateral and micro-radian angular control. The laser diodes are collimated using aspheric lenses and the first laser diode - Diode D of 785nm in figure~\ref{fig:setup} is aligned first by adjusting the 50:50 beam splitter. The optical assembly in the beam expander is fixed in the tube to get maximum power output of the collimated beam of a desired diameter of 3.6mm. The beam splitter is then fixed at the position using a UV curing glue, NOA 81. This glue is vacuum and space qualified. the viscosity of the glue is 300cps, which provides enough support to set the optical components at the right angular position without exerting unnecessary strain at the pick-up tool.
 The other diodes are collimated and then aligned similarly. The 4 alignment apertures in the bespoke aluminium casing help to ensure the power loss at every BS is not more than 3 dB and helps coarse positioning of the BS. The half waveplates and the filters are added at the end into the built-in grooves. The  casing also has a slot for a fibre-coupled laser in the optical axis of the pinhole on the other side, this is used to remove pointing error in the output by using a reference beam spot created by this fibre coupled laser.

 The eight laser diodes in the optical payload are controlled by sixteen laser drivers on a driver PCB in order to produce two different intensity pulses together with the vacuum state (so-called ``2-decoy-state'', or signal + decoy + vacuum) for each polarisation~\cite{lo2005decoy,lim2014concise}. One of these drivers is configured to trigger a laser diode with a higher current level than the other driver. A combination of low and high power laser driver actuation is then used to produce the different quantum states including signal state-lower intensity  and decoy state - high intensity. The source driver PCB includes eight Pulse Width Modulation (PWM) controlled current sink circuits that provide current bias for each of the laser diodes. This is necessary to stabilise voltage across the laser diodes and thus to achieve equally emitted optical power independent of trigger interval. Furthermore, monitor diode circuits are included for each of the eight laser diodes which provide voltage feedback, to estimate the optical power emitted at a given integration time interval. The rest of the circuit includes a Xilinx Field-programmable gate array (FPGA) with high speed low-voltage differential signaling (LVDS) lines that allow triggering laser drivers at their maximum allowable trigger frequency of 200 MHz, communication interfaces, programming interface, power regulators and storage devices.

\section{Results and Discussions}

\begin{figure} [!h]
   \begin{center}
   \begin{tabular}{c} 
   \includegraphics[width=\textwidth]{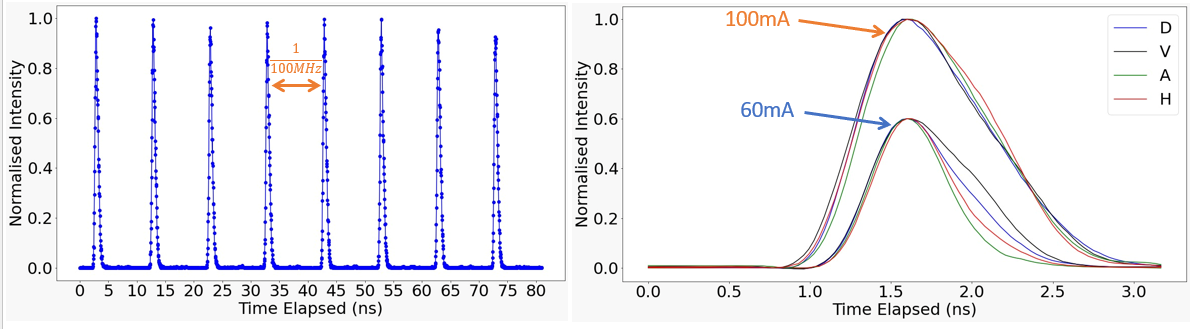}
	\end{tabular}
	\end{center}
   \caption[example] 
   { \label{fig:100mhz} (a) The left hand side graph shows the reconstructed screenshot from picoharp.  (b) The right hand graph shows the pulse widths of the four 785nm laser diodes at $60\,mA$ and  $100\,mA$ input current pulses for signal and decoy state. The average pulse width is observed to be $0.5\,ns$ and $0.9\,ns$}
   \end{figure}

The source is designed to pulse at 200 MHz on each laser diode, but due to the afterpulses at high repetition rates, the intensity of the optical pulses vary randomly. These afterpulses need to be characterised extensively in order to improve the pulsing repetition frequency. The pulses are uniform at the repetition rate of 100 MHz as shown in a reconstruction of data from the PicoHarp 300 Time-Correlated Single Photon Counting (TCSPC) in \ref{fig:100mhz}.  Temporal distinguishibility of the pulses from different laser diodes can pose a potential security risk. If the pulses have different systematic delays, the eavesdropper can estimate the state by the virtue of its arrival time without measuring its polarisation. Thus it is crucial to measure the pulse widths and triggering times of the laser diodes. The current version of the source module has no width discrepancy for different repetition rates but it has a shorter pulse width for the the decoy state. This loophole can be mitigated by characterising and equalising the pulse widths. Figure~\ref{fig:100mhz}b shows the data from the TCSPC with normalised intensities, The average pulse width full width half maximum (FWHM) for the decoy pulse is $500\,ps$ compared with $900\,ps$ for the signal state. This can be mitigated by either widening the pulse width of 60mA pulse to equalise, but that requires increasing the detection gate width as well, constituting to higher background noise which needs to be optimised acoording to the channel.

The different spatial modes can be used by the eavesdropper to extract some information about key using high resolution cameras~\cite{arteaga2022practical}, thus it is important to spatially mode filter the quantum beam to minimise any side channel attacks. Due to the space, weight and power restrictions we are using multimode, Febry-Perot TO-CAN packaged laser diodes which need to be mode filtered before transmission of quantum signal from the satellite. The mode filtering is done using the diffraction limited pinhole 5 $\mu m$ in a Keplerian beam expander as shown in Figure~\ref{fig:setup}. The space compatible lenses are fixed using a bespoke assembly that mode filters, expands and collimates the beam. The geometric divergence of the beam has to be $\le 17 \mu rad$ to meet the maximum input tolerance of the satellite output optics. In the current setup, the collimation is tested by done by measuring the beam waist at multiple distances up to $10\, m$ using a NIR camera to estimate the beam radius at different points. Since the single mode timing and synchronisation beacon signal~\cite{2021_Zhang} also passes through this assembly, the input beam diameters to the beam expander are optimised to reduce the loss due to the pinhole.
  
\begin{figure} [!h]
   \begin{center}
   \begin{tabular}{c} 
   \includegraphics[width=\textwidth]{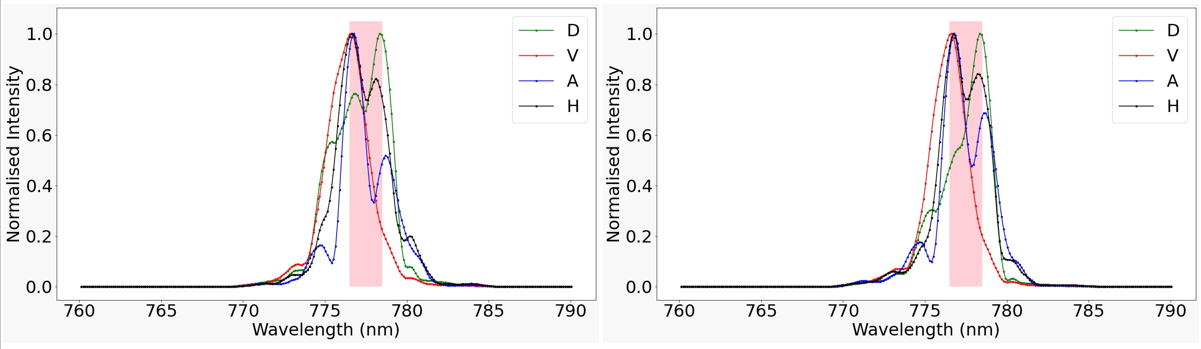}
	\end{tabular}
	\end{center}
   \caption[example] 
   { \label{fig:spectrum} (a) The left hand side graph shows the spectrum of the 4 laser diodes at $60\,mA$ input current pulsed at $100\,MHz$  and  (b) The right hand graph shows the spectrum at $100\,mA$ input current pulsed at $100\,MHz$. The range of the wavelength is within 5nm of 777.5nm central wavelength. The pink band shows the 2nm filter centered at 777.5 nm }
   \end{figure}

\begin{figure} [!h]
   \begin{center}
   \begin{tabular}{c} 
   \includegraphics[height= 5cm]{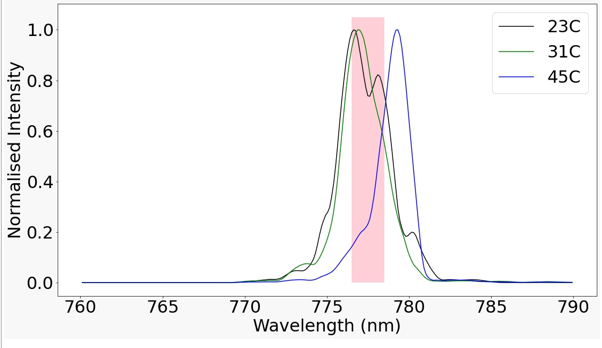}
	\end{tabular}
	\end{center}
   \caption[example] 
   { \label{fig:temperature} Temperature dependence of the spectrum of the 785 nm laser diodes, the pink band shows the 2 nm line width filter at 777.5 nm.}
   \end{figure}

It is crucial to ensure all the laser diodes emit at same wavelength to avoid the possibility of the eavesdropper estimating the polarisation state of the photon dependent on its wavelength. Thus, it is important to characterise the spectrum of all laser diodes under different driving conditions. Depending on the encoding pattern, the laser diodes will be transmitting the pulses at different repetition rates- 25 MHz, 50 MHz or 100 MHz. The spectrum of the laser diodes showed no shift with different repetition rates but since in decoy state operation, the diodes will be pulsing at different input currents, emitted wavelength shifts slightly, as shown in figure ~\ref{fig:spectrum}. The spectrum of the diodes also depends on the operational temperature, since the operating temperature of the satellite is meant to be maintained between $0-45^\circ$ C, the thermal test on the laser diode suggest minimal change in the wavelength as shown in figure~\ref{fig:temperature}. The wavelength range for the diodes at different temperatures and input currents lie within 5 nm around 777.5 nm.  Once the laser diodes are characterised in-situ, a 2 nm line width filter can be customised to the central wavelength and ensure the indistinguishable transmission spectrum as shown in a pink band on  ~\ref{fig:spectrum} and ~\ref{fig:temperature}. The characterisation data and power monitor measurement can be used to increase the pulse amplitude if necessary to equalise the intensity of the polarised pulses.

The Extinction ratio of the source module is the direct measure of quality of the source and purity of the polarization states it produces. Extinction ratio of this source module is measured using a Linear Polariser(LP). The LP is rotated between the source and a single photon detector to estimate the maximum and minimum counts at a $90^{\circ}$.The extinction values for horizontal,vertical, diagonal, and anti-diagonal polarizations are $0.61 \times 10^{-3} $, $0.35 \times 10^{-3}$, $1.3 \times 10^{-2}$, $1.8 \times 10^{-2}$ respectively. The extinction ratio contributes to the QBER with $0.817\pm 0.001\%$. The comparatively lower extinction ratio of the D/A bases is due to the birefringence of various optical components in the source module and can be corrected by using the Quarter,Half and Quarter waveplates in the receiver module. This would optimise the communication channel and reduce the overall QBER. 

\section{CONCLUSION}
\label{sec:sections}
 The paper highlights the major design steps and tests of the optical module of dual wavelength  WCP DS QKD source.  The source module assembly shows the various measures that are being taken in the design process to make it suitable to operate in a nanosatellite at LEO. The characterisation of the laser diodes at different frequencies, currents and temperature inspired the 2nm line width filter to mitigate any spectral side channel attacks. The spatial mode filtering of the output prevents any spatial side channel attacks. The decoy state protocol helps mitigate the photon splitting attacks and the temporal characterisation can be used to measure the timing side channel attacks. The Extinction ratio of the source will contribute by only $0.817 \pm 0.001\%$ to the intrinsic QBER of the QKD system. This system is characterised by Sidhu et al.\cite{sidhu2022finite} and can be used to generate a secret key for upto 40dB of channel loss between the ground station and the satellite. 
 Similar tests are conducted on the $808\,nm$ side of the source, but the results shown are just the $785\,nm $ half of the source. The two sources together will give the repetition rate of 200MHz.  The source is still under optimisation and further results from thermal, vibrational and shock testing will guide further iterations.


\acknowledgments 
J.S.   was   supported   by   the   Zutshi Smith Scholarship. D.K.L.O. is supported by the EPSRC Researcher in Residence programme at the Satellite Applications Catapult [EP/T517288/1] and the EPSRC International Network in Space Quantum Technologies [EP/W027011/1]. Part of the research leading to this work has been supported by the EPSRC Quantum Communications Hub funded by the EPSRC [grant ref. EP/T001011/1],  and UK Space Agency [NSTP3-FT2-065QSTP: Quantum Space Technology Payload, NSIP-N07 ROKS Discovery].

\bibliography{report} 
\bibliographystyle{spiebib} 

\end{document}